\documentclass[sigconf,screen]{acmart}

\definecolor{marroon}{HTML}{881c1c}
\definecolor{DarkBlue}{HTML}{00008B}
\definecolor{lgreen}{HTML}{e0f3db}
\definecolor{dpink}{HTML}{CD1076}
\definecolor{pink}{HTML}{FED2D2}
\definecolor{soothinggreen}{HTML}{4dac26}
\definecolor{darkred}{HTML}{8B0000}
\definecolor{dblue}{HTML}{104E8B}
\definecolor{violet}{HTML}{8A2BE2}
\definecolor{mscolor}{HTML}{01665e}
\definecolor{nmscolor}{HTML}{d8b365}
\definecolor{deepgrey}{HTML}{525252}
\definecolor{dslate}{HTML}{2F4F4F}
\definecolor{dolive}{HTML}{556B2F}
\definecolor{teal}{HTML}{388E8E}
\definecolor{gray}{HTML}{e8e8e8}

\usepackage{multirow,stfloats}
\usepackage{colortbl}
\usepackage{cleveref}

\newcommand{\basesize}{14 }
\newcommand{\obssize}{14 }

\newcommand{\numscorequantiles}{4 }
\newcommand{\ctrlgroupfactor}{3 }
\newcommand{\nstrata}{10 }
\newcommand{\minusersperstratum}{10 }
\newcommand{\smdthresh}{0.3}
\newcommand{\gcandidates}{$\mathcal{C}_{\text{gold}}$}
\newcommand{\scandidates}{$\mathcal{C}_{\text{score}}$}

\newcommand{\edit}[1]{#1}
\newcommand{\editprime}[1]{#1}

\AtBeginDocument{%
  }

\copyrightyear{2025}
\acmYear{2025}
\setcopyright{none}
\acmConference[CHI '25]{CHI Conference on Human Factors in Computing
Systems}{April 26-May 1, 2025}{Yokohama, Japan}
\acmBooktitle{CHI Conference on Human Factors in Computing Systems (CHI
'25), April 26-May 1, 2025, Yokohama,
Japan}
\acmDOI{10.1145/3706598.3713830}
\acmISBN{979-8-4007-1394-1/25/04}

\begin{document}

\title[A Quasi-Experimental Study of the Effects of Positive Feedback on Reddit]{Does Positive Reinforcement Work?: A Quasi-Experimental Study of the Effects of Positive Feedback on Reddit}

\author{Charlotte Lambert}
\email{cjl8@illinois.edu}
\orcid{0009-0002-8487-7485}
\affiliation{%
  \institution{University of Illinois Urbana-Champaign}
  \city{Urbana}
  \state{Illinois}
  \country{USA}
}

\author{Koustuv Saha}
\email{ksaha2@illinois.edu}
\orcid{0000-0002-8872-2934}
\affiliation{%
  \institution{University of Illinois Urbana-Champaign}
  \city{Urbana}
  \state{Illinois}
  \country{USA}
}

\author{Eshwar Chandrasekharan}
\email{eshwar@illinois.edu}
\orcid{0000-0002-7473-1418}
\affiliation{%
  \institution{University of Illinois Urbana-Champaign}
  \city{Urbana}
  \state{Illinois}
  \country{USA}
}

\renewcommand{\shortauthors}{Charlotte Lambert, Koustuv Saha, \& Eshwar Chandrasekharan}

\begin{abstract}
Social media platform design often incorporates explicit signals of positive feedback. Some moderators provide positive feedback with the goal of positive reinforcement, but are often unsure of their ability to actually influence user behavior. Despite its widespread use and theory touting positive feedback as crucial for user motivation, its effect on recipients is relatively unknown. This paper examines how positive feedback impacts Reddit users and evaluates its differential effects to understand who benefits most from receiving positive feedback. Through a causal inference study of 11M posts across 4 months, we find that users who received positive feedback made more frequent (2\% per day) and higher quality (57\% higher score; 2\% fewer removals per day) posts compared to a set of matched control users. Our findings highlight the need for platforms, communities, and moderators to expand their perspective on moderation and complement punitive approaches with positive reinforcement strategies to foster desirable behavior online.

\end{abstract}




\maketitle

\section{Introduction}

\begin{quote}
\small
``We heard you... awards are back!''-- Reddit Product Team\footnote{\label{note1}\url{https://reddit.com/r/reddit/comments/1css0ws/we_heard_you_awards_are_back/}}
\end{quote}

\noindent In September 2023, Reddit removed several positive feedback mechanisms, including awards and Reddit gold, from its interface\footnote{\url{https://reddit.com/r/reddit/comments/14ytp7s/reworking_awarding_changes_to_awards_coins_and/}} in an effort to create better ways to empower communities to reward content.
 Although \textit{gilding}---i.e., the ability to ``gild'' posts by donating a Reddit Premium subscription---is no longer a feature on Reddit, the platform brought back other forms of awards in May 2024\textsuperscript{\ref{note1}} based on feedback from its users.
These recent changes reflect both the platform's interest in supporting positive feedback and the users' desire to continue providing such feedback.

Positive feedback is a key feature of positive reinforcement, a long-studied behavior modification principle in the field of behavioral psychology shown to be effective in offline settings to encourage desired behavior through positive stimuli
~\cite{ferster_schedules_1957}.
In the online context, recent HCI research has revealed that some moderators of online communities consider positive reinforcement a part of their role~\cite{lambert_positive_2024}.
In particular, the system of rewards utilized in positive reinforcement involves the use of the available signals on the platform.
Importantly, most social media platforms include some form of positive feedback (e.g., Reddit upvotes, Facebook likes, etc.). 
In fact, the presence of positive feedback as an affordance is associated with a community consisting of higher-quality content~\cite{papakyriakopoulos_upvotes_2023}.
Furthermore, HCI researchers have established that positive feedback increases a user's motivation to contribute to a community~\cite{kraut_building_2011}. 
As a result, positive feedback may be vital both for users to want to participate and for platforms that rely on contributions to be successful. 
Despite the ubiquity of positive feedback across platforms and the established theories motivating its potential to have meaningful impacts offline and online,
there is a lack of empirical evidence supporting the effectiveness of receiving positive feedback on improving user-level outcomes.
Learning whether positive feedback can measurably change important outcomes can alter our perspective on moderation and motivate the need for platforms to support proactive approaches to moderation centered around reinforcement in conjunction with preexisting punitive approaches.

We ground our work in the taxonomy of moderation developed by \citet{grimmelmann_virtues_2015}.
The taxonomy highlights openness through community participation in moderation as an important goal that encourages democracy.
\citet{grimmelmann_virtues_2015} also mentions two key moderation actions, organization and norm-setting, to help shape the content in feeds and create shared norms between community-members, respectively.

Distributed moderation has been theorized to be effective~\cite{grimmelmann_virtues_2015} and has been found effective at enforcing norms in practice~\cite{lampe_crowdsourcing_2014}. 
While Reddit is ostensibly moderated by centralized moderation teams, the use of widespread positive feedback inherently creates a democratic system of distributed moderation that enables users to voice opinions on what content should be encouraged, aligning with the moderation goal of openness~\cite{grimmelmann_virtues_2015}.
We need to understand the cause-and-effect relationships between these positive-feedback-based distributed moderation practices and the other goals of moderation (i.e., reorganizing communities and facilitating norm-setting) to inform design solutions that can leverage the power of positive feedback to promote healthier communities.

\begin{figure*}
    \centering
    \includegraphics[width=0.89\linewidth]{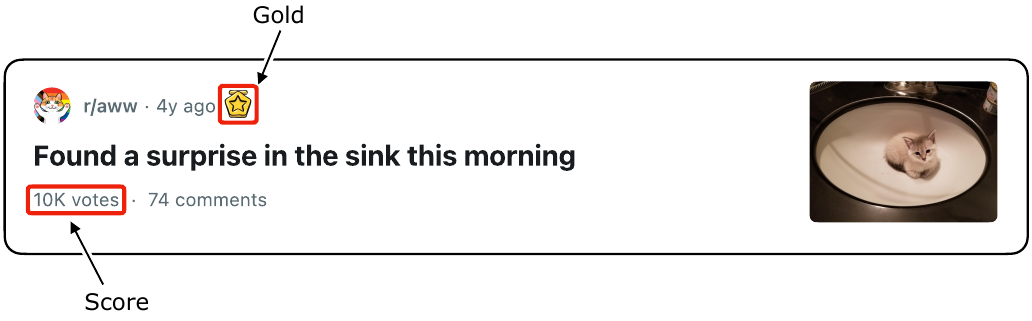}
    \caption{This is an example post from r/aww that received gold, represented by a small gold icon above the title. The post's score is reported under its title. This post's score was in the top 25th percentile for r/aww in the month it was posted.}
    \Description{Reddit post from r/aww that contains a picture of a small cat in a sink. The post received one gold and a score of 10K.}
    \label{fig:treatment_example}
\end{figure*}

To explore distributed moderation through positive feedback, we focus on Reddit gold and upvotes, two forms of positive feedback available to all community members. \autoref{fig:treatment_example} provides an example post that received both forms of feedback.
Gold was a beloved~\cite{weatherbed_2024_reddit} but sparsely-used, paid feature while upvotes are free to give out and used in all sub-communities on Reddit.
Prior research confirms that moderators utilize both gold and upvotes as mechanisms for positive reinforcement~\cite{lambert_positive_2024}, however both forms of feedback are given anonymously, so we cannot evaluate their efficacy as moderator-specific feedback and instead explore whether they are capable of accomplishing the goals of moderation when given by the community more generally (i.e., distributed). 
\subsection{Research Questions}

The posting quality, posting frequency, and norm-adherence of users are important to 
communities~\cite{weld_making_2024}, 
platforms~\cite{kraut_building_2011}, and moderators~\cite{lambert_positive_2024} alike. 
Thus, we ask whether positive reinforcement through distributed moderation is capable of influencing these important user-level outcomes. More specifically, we pose the following research questions:

\begin{enumerate}
\item[\textbf{RQ1:}] What is the effect of positive feedback on recipients?
\begin{enumerate}
    \item How does receiving gold or being highly upvoted impact users' future behavior in a community?
    \item How long do the effects of positive feedback on user behavior last?
\end{enumerate}
\item[\textbf{RQ2:}] How do the effects of positive feedback vary across different types of recipients?
\begin{enumerate}
\item What types of users experience stronger effects of receiving positive feedback?
\item What types of communities experience stronger average effects of receiving positive feedback?
\end{enumerate}

\end{enumerate}

\subsection{Summary of Contributions}

\subsubsection{Methods.}
With these research questions in mind, we employ causal inference methods to estimate the causal effects of receiving positive feedback (i.e., our treatments). 
\edit{In particular, we draw on the potential outcomes framework~\cite{rubin_causal_2005, imbens_causal_2015} to conduct stratified matching followed by difference-in-difference analyses. To do so,}
we first form a group of treatment posts which received a large quantity of upvotes and/or received gold. We then form a control group of similar posts from the same communities that did not receive such forms of positive feedback. With these two groups, we perform \edit{stratified matching to match users with similar propensities of being treated. Then, we conduct} a difference-in-differences analysis to understand the effect of receiving positive feedback on a user’s future behavior in their community and compare pre- and post-treatment behavior to understand how long these effects persist. 

We then explore the differential effects of our treatments through regressions on the individual treatment effects to 
identify which types of users and communities experience stronger effects of receiving positive feedback. 

\subsubsection{Findings.}
We find that users who receive gold or high score on a post 
receive more positive feedback on future posts compared to if they had never received the positive feedback. Highly-upvoted users also experience an increase in their posting rate and a decrease in their rate of removals. Additionally, we observe that users who received stronger treatment (i.e., more gold or upvotes) experience larger increases in future positive feedback received. Importantly, Reddit newcomers experience larger increases in posting frequency and larger decreases in removal rates after being highly upvoted.

\subsubsection{Implications.}

This work contributes to HCI research by quantifying the power user interface elements have on influencing user behavior. Specifically, some researchers have developed their own systems to encourage certain user behaviors~\cite{im_synthesized_2020, chang_thread_2022, yen_storychat_2023} and others have explored the effect of existing interface features, such as badges~\cite{anderson_steering_2013} and reputation systems~\cite{adler_content-driven_2007, resnick_reputation_2000}, on users.
We build on this work by highlighting upvotes, a signal nearly universally available in some form across social media platforms, and by gaining a deeper understanding of how users across platforms may benefit from the use of positive feedback. 
We also advance moderation research by expanding our understanding of how community-driven moderation can function in conjunction with positive reinforcement strategies. 
Based on our findings, we make design recommendations for platforms and HCI researchers to facilitate the use of positive feedback signals for encouraging desirable behavior and easing norm acquisition.
Furthermore, we identify specific populations (e.g., newcomers, users who have not recently experienced positive feedback) that may gain the most from receiving positive feedback. 
We also highlight the benefits of using positive feedback for platforms and moderators and call on designers to create explicit interface features to support this form of proactive moderation.
Finally, we highlight ways moderators can work with the current system of positive feedback to guide users towards norm-abiding and high-quality content.
\section{Background}

This section provides an overview of relevant background literature related to behavioral psychology, distributed moderation theory, the effect of interface signals, and gift-giving in online communities.

\subsection{Principles in Behavioral Psychology}

Early research in behavioral psychology involved exploring the effect of positive consequences on encouraging behavior, firstly through Thorndike's law of effect~\cite{gray_psychology_2010}. This principle states that people repeat behavior with satisfying consequences and avoid behavior with unpleasant consequences. Thorndike's principles have been studied in experimental setups~\cite{athalye_evidence_2018}, but subsequent research building on the law of effect has been more thoroughly validated. 

More specifically, building on the law of effect, B. F. Skinner developed the idea of positive and negative reinforcement and punishment in his work on operant conditioning~\cite{skinner_operant_1963, ferster_schedules_1957, skinner_science_1965, skinner_recent_1989}. 
Prior moderation research has often focused on the use of punitive techniques such as content removals, bans, and quarantines, but our work centers around \textit{positive reinforcement}, the introduction of a positive stimulus to encourage behavior, to evaluate whether it can encourage users online to repeat the behavior that led to the positive outcome. The use of positive reinforcement has been validated in offline settings like workplaces~\cite{bradler_employee_2016, perryer_enhancing_2016,armstrong_gamification_2018, wei_impact_2014}, education~\cite{lysakowski_classroom_1981, anderson_engaging_2014, boyd_direct_1981}, parenting~\cite{steinberg_impact_1992,dinkmeyer_parents_1989}, and athletic training~\cite{wiese_sport_1991, fayyaz_view_2021, weinberg_effect_1990}.

One specific type of reinforcing feedback that has been studied is praise. \citet{kazdin_history_1978} find that praising desired behavior increases the likelihood that the behavior will be repeated. On Reddit, praise happens through upvoting and awarding signals. Prior work found that some moderators specifically provide positive feedback like gold and upvotes to reinforce behavior and others specifically aimed to praise the author~\cite{lambert_positive_2024}.

Additional research has identified another type of reinforcement: vicarious. The idea of vicarious reinforcement is that bystanders of positive reinforcement may be vicariously influenced to repeat the behavior being reinforced~\cite{bandura_conditions_1971, kazdin_effect_1973}. 
\citet{jhaver_bystanders_2024} explore the effect of exposure to post-removal explanations on bystanders, essentially vicarious punishment. Though they found that the exposure did not encourage bystanders to learn the norms of the community, prior work in in social computing shows that vicarious reinforcement through highlighting examples of norm-adhering behavior in a community may help with further norm-adherence~\cite{kiesler_regulating_2012}. 
\citet{seering_shaping_2017} showed that users on Twitch imitated observed behaviors, especially from high status users.
Furthermore, some users believe providing positive feedback can increase a post's visibility in the feed~\cite{lambert_positive_2024}, effectively highlighting the content for the rest of the community. 
This exposure to posts approved of by the community may enable community norm acquisition by newcomers and other bystanders through vicarious reinforcement. 

\citet{miltenberger_behavior_2016} further explored the factors that may impact the effectiveness of positive reinforcement strategies in practice and presents three notable factors: contingency, satiation, and immediacy. Contingency states that a positive stimulus will have the strongest effect when it is only presented in response to the behavior being reinforced and at no other time. Satiation is the idea that a positive stimulus will have a diminished effect on the recipient if they have been too exposed to the stimulus. \citet{wang_highlighting_2021} demonstrate the idea of satiation in their work which reveals that receiving a New York Times Pick badge, a form of positive reinforcement, has the largest effect on first-time recipients' future comment quality and diminishing effects for more frequent recipients. Finally, the concept of immediacy indicates that the reinforcing stimulus will be more effective when received more quickly.
\citet{choi_creator_2024} explored the idea of immediacy in the context of YouTube creator hearts and found that videos received more engagement when creator hearts are given sooner after the video was published.

\textit{Our work builds on the theory of operant conditioning and prior work demonstrating the effectiveness of positive reinforcement in offline settings to understand community-driven moderation on Reddit.}

\subsection{Distributed Moderation}
\edit{Reddit's moderation system exhibits qualities of both centralized and distributed moderation systems. There are some platform-level moderation structures (e.g., admins\footnote{\url{https://redditinc.com/policies/content-policy}}), but sub-communities create their own sets of rules~\cite{reddy_evolution_2023} and most moderation falls to volunteer moderation teams within each community. While these volunteers themselves are Reddit users, they still form centralized moderation structures within communities.}
These moderation teams are crucial to user safety, but the \edit{non-moderating users} play a similarly important role in moderation. Reddit enables users to give various forms of feedback to contributions, such as upvotes, downvotes, and awards. These signals are then taken into consideration for the sorting of community feeds. In this way, users have the collective power to highlight content they approve of and hide content they do not. In practice, this is an example of distributed moderation, a system through which members of a community are responsible for making moderation decisions. 

There is extensive research on Reddit moderation, including identifying abusive behavior~\cite{chandrasekharan_bag_2017, habib_act_2019, urbaniak_namespotting_2022, almerekhi_investigating_2020,bagga_are_2021}, evaluating the efficacy of moderation strategies~\cite{chandrasekharan_quarantined_2022, chandrasekharan_you_2017, srinivasan_content_2019,jhaver_does_2019, matias_preventing_2019}, and characterizing moderator experiences~\cite{schaffner_community_2024, kiene_who_2020}. However, prior work does not often explore the power of Reddit's distributed moderation through user feedback, though some Reddit moderators themselves have stated that users should utilize upvotes and downvotes on the platform to self-regulate, demonstrating support for distributed moderation from active participants in Reddit's centralized moderation~\cite{lambert_positive_2024}.

Prior work into distributed moderation often focuses on the platform Slashdot. \edit{\citet{lampe_follow_2005} specifically highlight the effect of positive and negative feedback given to an author's first contribution on their future participation on Slashdot. Additionally,} \citet{lampe_slash_2004} and \citet{lampe_crowdsourcing_2014} show the power distributed moderation on Slashdot has to identify high and low quality contributions, while also reporting some of the challenges associated with the approach. 
\citet{jiang_trade-off-centered_2023} and \citet{grimmelmann_virtues_2015} similarly identify trade-offs between various moderation practices, including comparisons of centralized and distributed moderation. \citet{jiang_trade-off-centered_2023} deem distributed moderation a more democratic process as it better reflects community-wide opinions.

In this work, we build on prior work presenting the advantages of distributed moderation and demonstrating the support for such an approach, even from members of centralized moderation teams, to understand its effect in practice. \textit{We focus specifically on forms of positive feedback to advance our understanding of whether distributed moderation on Reddit is an effective form of community-driven positive reinforcement.}

\subsection{Guiding User Behavior Through Interface Signals}

Platform designers have the power to determine what users can do in their communities by choosing to include or exclude various interface elements. For example, some platforms utilize badges~\cite{anderson_steering_2013}, gamification mechanisms~\cite{papoutsoglou_modeling_2020, hamari_does_2014, sailer_how_2017}, or reputation systems~\cite{resnick_reputation_2000, resnick_value_2006, adler_content-driven_2007}, which have all been shown to encourage participation.
Prior work has also shown that users can be encouraged to make higher-quality contributions through the use of specific highlighting mechanisms~\cite{wang_highlighting_2021}. Similarly, posts on Reddit receive one upvote by default, an instance of example-setting that may encourage other users to follow suit. 
\edit{Upvotes and other explicit signals of positive feedback on Reddit are examples of formal feedback, which may be more effective at encouraging norm awareness and abidance than informal forms of feedback~\cite{kiesler_regulating_2012}.}
Finally, other research has shown that prominently displaying community rules and guidelines encourages adherence, especially from newcomers~\cite{matias_preventing_2019}. This practice is used across Reddit and Twitch.\footnote{\url{https://safety.twitch.tv/s/article/Chat-Tools}}

Social computing researchers have developed new tools and signals on platforms to encourage certain behaviors. \citet{im_synthesized_2020}, for example, introduced a signal to the Twitter interface to inform users about an account's history with toxicity and misinformation. 
\citet{chang_thread_2022} introduce an intervention for Reddit users engaging with inflammatory content to encourage more reflection about whether they should contribute to uncivil conversations.
Similarly, \citet{yen_storychat_2023} build a tool to visualize the amount of negativity in live-streaming chatrooms to encourage viewers to proactively moderate the space by contributing prosocially. 

Specific user behaviors can also be encouraged through more high-level considerations. For example, platform designers determine whether users are able to remain anonymous in their contributions which can influence users' willingness to contribute certain types of content. More specifically,
prior work has found that anonymity leads to more sensitive self-disclosures~\cite{peddinti_internet_2014, clark-gordon_anonymity_2019, ma_anonymity_2016} and more thoughtful feedback~\cite{choudhury_mental_2014}. However, in some communities, this may come at the cost of lower-quality contributions ~\cite{omernick_impact_2013} or more disruptions~\cite{donath_identity_1998}.

Similarly, platforms can consider social translucence~\cite{gilbert_designing_2012, erickson_social_2000}, the concept of emphasizing the visibility of social information in platform design, when making decisions. 
For example, X (formerly Twitter) has recently changed the visibility of users' liked posts so others can no longer see who liked another user's post, nor can users see the posts other users have liked.\footnote{\url{https://x.com/XEng/status/1800634371906380067}} Similarly, Instagram now provides the option to make the like count on posts private.\footnote{\url{https://about.instagram.com/blog/announcements/giving-people-more-control}} Ostensibly these changes are motivated by user privacy or maintaining user well-being, but there is disagreement in prior work regarding whether like counts on Instagram are responsible for negative effects on user well-being~\cite{wallace_hiding_2021}. Additionally, there are negative side effects of these adjustments on social translucence because they mask vital network information in a way that may hinder interactions between users.

However, certain interface features can enable users to influence their communities. Reddit upvotes, for example, are a widespread mechanism used across the platform and are controlled entirely by Reddit users.
There has been prior research in understanding the role of upvotes on Reddit~\cite{morrison_here_2013, lambert_investigating_2024}, including in spaces like advice-giving~\cite{lu_audience_2024} and political communities~\cite{papakyriakopoulos_upvotes_2023,carman_manipulating_2018}. Prior work has shown that interfaces that allow upvotes typically have higher-quality content~\cite{papakyriakopoulos_upvotes_2023}. Additionally, \citet{carman_manipulating_2018} manipulated the number of upvotes given to posts in political and non-political subreddits and found that posts with artificially-inflated upvote counts experienced increases in their final upvote and reply counts, demonstrating the power of upvoting to increase engagement and positive feedback. \textit{We extend this research to understand the specific influence positive feedback has on the recipients themselves.}

\subsection{Gift-Giving and Reciprocity}

Social computing literature highlights gift-giving as a mechanism that builds a sense of reciprocity from gift-givers. Essentially, users expect to receive similar treatment from their communities as they give~\cite{kollock_economies_1999}. Other work shows that receiving a gift encourages users to give similar gifts in the future~\cite{kizilcec_social_2018}, highlighting reciprocity for the recipient. Importantly, gift-giving is a means to build reputation which social computing literature touts as a clear signal of desired behaviors~\cite{lampe_role_2012}, an incentive for good behavior~\cite{resnick_reputation_2000}, and a tool to build trust and engagement among users~\cite{resnick_value_2006}. 

In Reddit communities, positive feedback mechanisms are staples of gift-giving, and if they abide by the principles of reciprocity, recipients of positive feedback may be motivated to provide similar feedback in the future.
This highlights the importance of understanding user-level outcomes of gift-giving through positive feedback.
\textit{Our work explores the user-level outcomes in response gift-giving in the form of Reddit gold and upvotes.}

\section{Study Design and Data}
\begin{figure*}[t]
    \centering
    \includegraphics[width=0.95\linewidth]{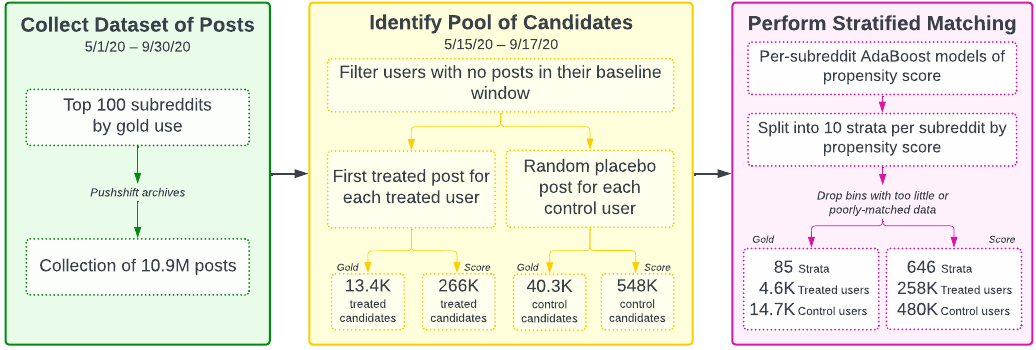}
    \caption{Pipeline from data collection to forming matched treated and control samples used in our causal inference methodology.}
    \Description{Pipeline diagram of our data collection and analysis, split into three boxes. The first box indicates the collection of the dataset from Pushshift's archives of the top 100 subreddits that gild most frequently, resulting in 10.9M posts. The second box shows how we identify a pool of candidates from this large initial sample. After various filters we have 13.4K treated and 40.3K control candidates for the gold treatment. For the score treatment, we have 266K treated and 548K control candidates. The final box depicts our process for performing stratified matching and eventually results in 85 strata made of 4.6K treated and 14.7K control users for the gold treatment. The score treatment is left with 646 strata consisting of 258K treated and 480K control users.}
    \label{fig:pipeline}
\end{figure*}
This section describes our data-collection process and the causal-inference approach we adopted drawing on the potential outcomes framework~\cite{rosenbaum_assessing_1983, rubin_causal_2005} used in prior social computing research~\cite{jhaver_bystanders_2024, chandrasekharan_you_2017, saha_prevalence_2019}.
\edit{Our goal is to examine the causal effects of receiving positive feedback (i.e., the treatment) compared to what would have happened if the treatment was not administered (i.e., the \textit{counterfactual}).
However, it is not possible to find a post that simultaneously did and did not receive positive feedback, so a \textit{true} counterfactual is missing or unobserved.
The potential outcomes framework allows us to estimate the missing counterfactual for each treated post based on the outcomes of similar, observed posts that were not treated (i.e., did not receive positive feedback).}
\edit{We estimate the counterfactual by} identifying a treatment group (posts that receive positive feedback) and a \edit{similar control group} (posts that did not receive positive feedback). We identify similar posts from each group to create matches, allowing us to compare the trajectory of similar users who differ only in whether they received our treatment (i.e., positive feedback).
The full pipeline of this process is visualized in \autoref{fig:pipeline} along with descriptive statistics about the dataset.

\subsection{Treatments: Posts That Received Positive Feedback}

There are many forms of feedback available on Reddit. We select two to focus on in this work: \textit{gold} and \textit{score}. \autoref{fig:treatment_example} visualizes both forms of feedback in a real Reddit post.

Gold was an award-like signal consisting of gold icons that were awarded to posts. Gold and other awards were discontinued by Reddit in 2023,\footnote{\url{https://www.reddit.com/r/reddit/comments/14ytp7s/reworking_awarding_changes_to_awards_coins_and/}}
however gold was considered beloved by some users on Reddit~\cite{weatherbed_2024_reddit} and many users were disappointed that they were not among the features brought back to the platform in May 2024.
Thus, we believe there is still value in exploring how gold affected the recipients to understand whether Reddit removed a mechanism that had positive impacts on user behavior and experience. 

Because gold was a paid feature, it is a potentially strong signal of quality given the rarity with which it was given out. However, the real-world cost of gilding means that gold may not be entirely representative of all opinions or values from a community. To account for this, we select score (i.e., aggregated upvotes and downvotes) as our second treatment. Upvotes are the most widespread signal of approval on Reddit, utilized by every community. To contrast gold, upvotes are free to give out, therefore there is no limit on how many a user can hand out, neither explicitly imposed by the platform nor implicitly imposed by cost. We note that some moderators believe that upvotes are not necessarily a reliable indicator of quality, specifically in the context of communities like r/AskHistorians that value input from experts~\cite{jiang_trade-off-centered_2023}. However, these moderators are not claiming that upvotes cannot reliably inform us what a community likes. Thus, we recognize this limitation of our work and do not claim that the posts treated with upvotes contain accurate information or even content that moderators themselves would like to encourage. 
Furthermore, we rely on the concept of contingency~\cite{miltenberger_behavior_2016} 
by assuming that receiving the treatments is contingent upon posting something deserving of reinforcement within the community. Thus, we trust community signals of approval and do not evaluate how deserving the treated posts are.
This mimics approaches of prior work which focus only on the outcomes of treatment and do not consider the precursors leading to the treatment being applied~\cite{jhaver_does_2019,jhaver_bystanders_2024}.

\subsection{Identify Pool of Candidates}
Following approaches in prior work~\cite{jhaver_bystanders_2024, chandrasekharan_you_2017, kiciman_using_2018}, we use these two forms of treatment to identify treated and control candidates.
To begin, we use Pushshift~\cite{baumgartner_pushshift_2020} archives of Reddit data from May 1, 2020 to September 30, 2020 (our study period).
This research did not involve any participation from users and all data was publicly available on Reddit, thus our work did not require approval from an institutional review board. We followed best practices to maintain anonymity in our data (e.g., looking at data only in aggregate) and ensure user privacy.

Since upvotes are used universally and gold was not utilized by all communities, we construct a sample from the 100 subreddits which awarded the most gold in the study period.
We form  distinct pools of treatment and control posts, \gcandidates{}  and \scandidates{}, for each treatment,
made within our \textit{treatment window}: May 15, 2020 to September 17, 2020. This smaller window allows us to study user behavior in the \basesize days before and after treatment.
Posts which received at least one gold are considered treated candidates in \gcandidates{}. Posts that receive no gold are added to \gcandidates{}  as control candidates. 

To construct \scandidates{}, we need to identify some threshold above which a post is considered to be highly upvoted. To accomplish this, we split the distribution of scores for all posts in each subreddit into \numscorequantiles quantiles per month. The highest quantile consists of the highest-scored posts which are added to \scandidates{}  as treated candidates. Posts in the bottom two quantiles comprise the control candidates in \scandidates. We drop all posts in the second highest quantile to provide a buffer between control scores and treated scores, ensuring that our treatment is strongly defined. 

\begin{figure*}
    \centering
    \includegraphics[width=0.85\linewidth]{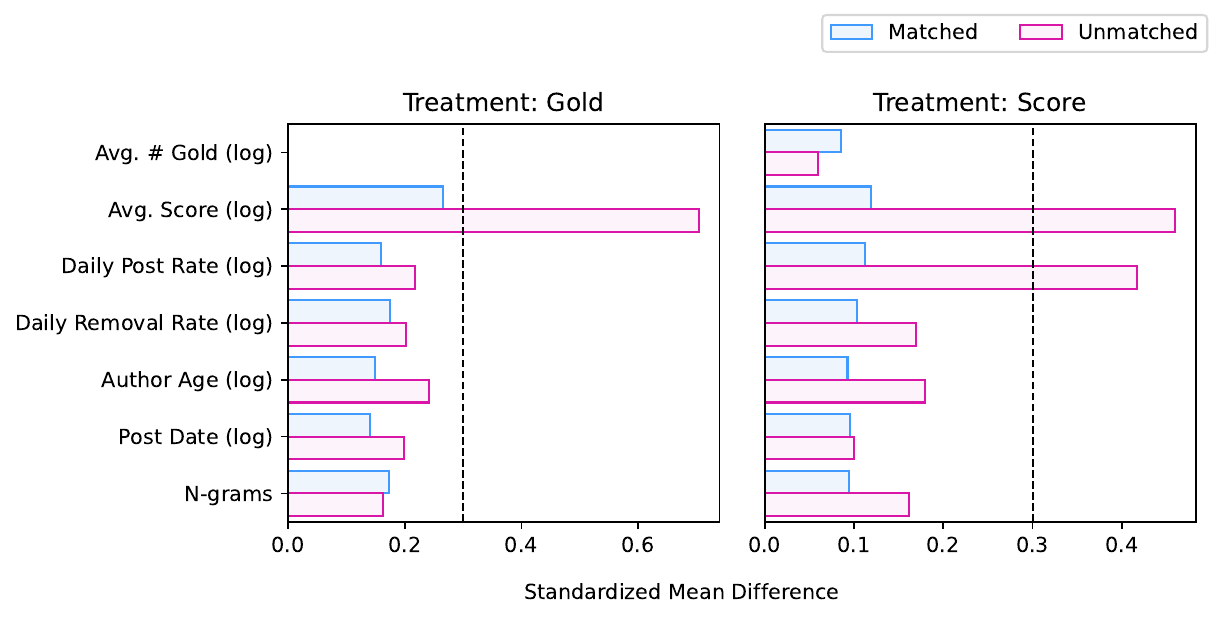}
    \caption{Standardized mean differences (SMD) for each covariate averaged across all strata. We report both SMDs for matched data and unmatched data for comparison. The figure shows that matching improved the SMD for most covariates and that the SMD for all matched data is below the threshold \smdthresh{}, indicating high-quality matches.}
    \Description{A plot with two sub-plots. The leftmost plot reports the average Standardized Mean Difference (SMD) of each covariate for the matched data using the gold treatment. The plot shows that the unmatched data has higher SMD values than the matched data for all covariates except the N-grams. A dotted line at the SMD value of 0.3 shows that all matched data has SMD values below our threshold of 0.3. The rightmost plot visualizes the same measures for the matched data using the score treatment. All covariates except the average number of gold has a lower SMD after matching, however no matched data surpasses the line at 0.3 SMD.}
    \label{fig:SMD}
\end{figure*}

For each treated and control candidate in \gcandidates{}  and \scandidates{}, we ensure that the author of the post has made at least one post in the \basesize days prior, none of which received the respective treatment. This provides us with baseline activity levels and avoids users receiving multiple treatments in a short span of time.
In this setup, it is possible for one user to appear in both candidate pools multiple times, a fluid interpretation of our treatments since the start of our study period is arbitrary so we do not know which users receive treatment before the start of our analysis.

For both treatments, we sample \ctrlgroupfactor times as many control users from each subreddit as there are treated users to balance the groups while still providing sufficient control candidates for matching. \autoref{fig:pipeline} reports summary statistics about the final pools of candidates, \gcandidates{}  and \scandidates.

For each candidate in \gcandidates{} and \scandidates{}, we dynamically define three events of interest:
\begin{itemize}
    \item \textbf{Baseline window:} the posting activity of the post's author in the \basesize days prior to treatment or placebo.
    \item \textbf{Treatment/placebo time:} the creation time of the post that received the treatment or placebo.
    \item \textbf{Observation window}: the posting activity of the post's author in the \obssize days after treatment/placebo time. 
\end{itemize}

\edit{We note that our analysis is limited by 
the fact that the precise time at which positive feedback is given is not publicly accessible, so we do not have confirmation that the recipients are aware that the treatment has been applied.
We use the creation time of the treated post as a proxy for treatment time which may raise issues with immediacy~\cite{miltenberger_behavior_2016}, in that authors are likely not treated immediately after contributing and they may not realize they received the treatment until even later. We could have further restricted the treatment pool to authors who also made comments on their gilded post, however it is not guaranteed that the gold was given prior to their comment. Thus, we consider our analysis to be a conservative estimate of the effect of positive feedback on users.}

\subsection{Stratified Matching}
\label{matching}

We want to minimize the differences in pre-treatment author and environmental characteristics between the control and treatment groups so observed effects can be attributed to the treatments. For each treated user, our goal is to identify similar control users from the same community based on \edit{treatment or placebo post content} and pre-treatment posting behavior and content quality.

We perform within-subreddit stratified matching on propensity scores in each candidate pool. Stratified propensity score matching has been used in prior work~\cite{kiciman_using_2018, saha_causal_2020, saha_social_2018, verma_examining_2022, yuan_mental_2023, olteanu_distilling_2017} to minimize biases of propensity score matching identified by \citet{king_why_2019}. 
This approach allows us to handle the bias-variance trade-off by striking a balance between too biased (one-to-one matched) and too variant (unmatched) data comparisons so we can isolate and examine the effects of treatment within each stratum~\cite{jhaver_bystanders_2024,saha_advertiming_2021}.
For each subreddit, we train an AdaBoost classifier which outputs a propensity score given the following covariates (log-scaled to achieve more normal distributions) as independent variables:

\begin{itemize}
        \item Daily posting rate in the baseline window.
        \item Average score on posts in the baseline period.
        \item Treatment/placebo time.
        \item Author's age (i.e., days since account creation) at treatment/ placebo time.
        \item Daily removal rate in the baseline period.
        \item Frequency of 100 bigrams from the \edit{text contained in the treatment/placebo post (i.e., titles and post bodies). We use the 100 most common bigrams appearing in all of the posts in \gcandidates{} or \scandidates{}.}
    \end{itemize}
When considering score as our treatment, we also include the following covariate:
\begin{itemize}
    \item Average number of gold received per post in the baseline period.
\end{itemize}
This variable does not apply to authors of posts in \gcandidates{} because, by definition, they must not have been gilded in their baseline period.

The resulting propensity scores reflect how likely a user is to belong to the treatment group based on the covariates. 
\edit{These covariates allow us to match treatment and control users that have similar ``quality'' posts in the pre-treatment window, so the authors should be similarly good writers before the treatment occurs. After matching, any effects observed after the treatment is applied can be attributed to the treatment itself.}
Within each subreddit, we bin the propensity scores into \nstrata strata using quantiles, filter out the strata with fewer than \minusersperstratum users, and ensure that each stratum contains both treated and control users.

Each resulting stratum represents a group of users with similar likelihood of being treated, thus the strata are a set of many-to-many matches for each subreddit such that every treated user in a stratum is matched with all control users in the stratum. We assign each stratum a unique identifier to be used in subsequent analysis. We refer to the stratum identifier of author $a$ as $s_a$.

To measure the quality of our matches, we compute the Standardized Mean Difference (SMD) between the treatment and control samples of each stratum for each covariate. We use an upper bound of \smdthresh{} to  indicate good match quality overall~\cite{kiciman_using_2018,kampenes_systematic_2007}. 
We drop all strata with a mean SMD over all covariates that exceeds \smdthresh{} to ensure that we retain only well-matched treatment and control posts and obtain an unbiased estimate of the treatment effects.
\autoref{fig:SMD} visualizes the SMD values for each of our covariates averaged across the remaining strata (i.e., \textit{matched} samples) and across the original pool of candidates (i.e., \textit{unmatched} samples) to show that matching improved the SMDs for our covariates. 
We note that the SMDs for two of the covariates were slightly lower before matching,
however the mean matched SMDs are still below our threshold of \smdthresh{}. We report descriptive statistics about the final matches in  \autoref{fig:pipeline}.

\subsection{Outcomes}
With this dataset of well-matched treatment and control posts, we examine the effects of receiving positive feedback on \editprime{four} major user-level themes: reception from the community, \editprime{content,} engagement, and norm-adherence. 
To measure community reception, we look at the effect of our treatments on each user's future \textit{average number of gold} and future \textit{average score} per day. These outcomes capture whether receiving positive feedback encourages users to post more positive-feedback-worthy content in the future.
\editprime{To understand the effect of positive feedback on content, we measure the \textit{text difference} between the treatment or placebo post and the recipient's future contributions. This difference is measured by computing sentence embeddings~\cite{reimers_sentence-BERT_2019} for the text in each post's title and body and measuring the cosine distance to the text of their treated or placebo post.}
For engagement, prior work describes positive feedback as a mechanism capable of increasing user motivation to participate~\citet{kraut_building_2011}. We use \textit{daily posting rate} as a proxy for user motivation to participate.
Finally, to capture the impact of treatment on norm-adherence, we utilize moderation decisions by each community's centralized moderation team through each user's \textit{daily removal rate}.

For each post in our matched sample, we collect all other posts by the same author in the baseline and observation windows. This gives us a set of 309K posts for our gold treatment and 5.4M posts for our score treatment. We calculate daily values for each of our \editprime{five} outcomes, taking care to exclude the treated or placebo post to not skew any averages.
This results in 129K daily aggregates for the gold treatment and 2.6M daily aggregates for the score treatment.
\begin{figure*}[t]
    \centering
    \includegraphics[width=0.99\textwidth]{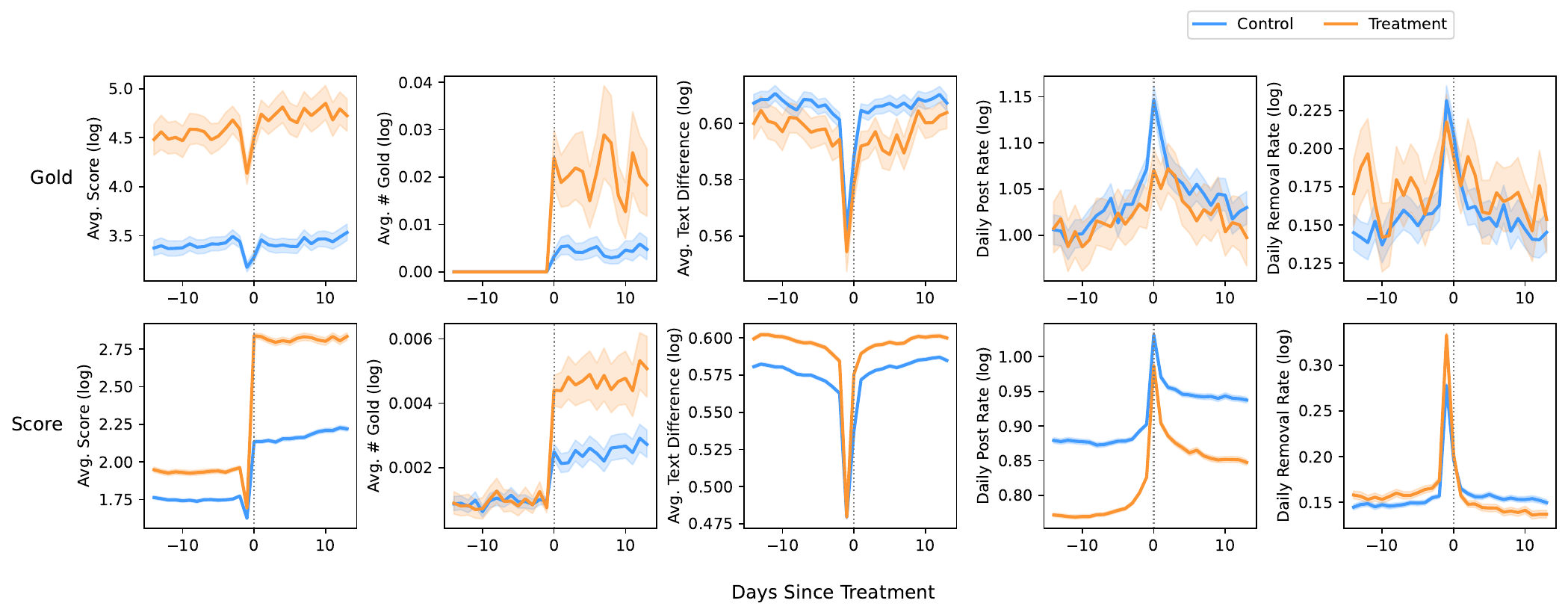}
    \caption{Visualization of each outcome over the baseline and observation windows (excluding treatment/placebo posts themselves). These plots are used to validate the parallel trends assumption needed for difference-in-differences analysis.}
    \Description{This figure has ten subplots. The top row of five plots visualize each outcome over time for the gilded posts and their matches. The bottom five subplots visualize the same outcomes for the highly-upvoted posts and their matches. The plots start at the beginning of the baseline period (day -14) and extend to the end of the observation window (day 13). These plots show that most outcomes are parallel prior to treatment between the treatment and control groups, but the treatment groups experience different effects after treatment compared to the control groups.}
    \label{fig:pta}
\end{figure*}
\section{RQ1: What is the effect of positive feedback on recipients?}

Using our matched treatment and control samples, we perform analyses to understand the causal impact of receiving positive feedback on user behavior and how long those effects last.

\subsection{RQ1a: How does receiving gold or being highly upvoted impact users' future behavior in a community?}

To understand how the treatments affect users (RQ1a), we conduct a Difference-in-Differences (DiD) analysis similar to analyses done in prior work~\cite{chandrasekharan_you_2017, abadie_semiparametric_2005,grinberg_changes_2016}.
The DiD analysis compares the difference in treated user's behavior before and after receiving treatment against the difference the control group experiences over time.
\edit{Like any causal inference study using observational data, there are limitations to our approach~\cite{weld_adjusting_2022}. Our matching process accounts for measured confounds in our data (i.e., the covariates) and the DiD analysis captures time-invariant unmeasured confounds,} for example, users in small communities experiencing a surge in subscribers may receive higher score on their posts across the board.  \edit{ It is possible, however, that external events impact how a community reacts to content in ways not captured by our observational data, leaving some time-variant unmeasured confounds unaccounted for. Future work can employ true experiments to account for such confounds and extend our understanding of the impact of positive feedback.}

To estimate causal effects through DiD analysis, the data must satisfy the \textit{parallel trends assumption}~\cite{did_columbia_2016}. This means the difference between the outcomes for the treatment and control groups must be constant over time, prior to treatment. Similar to prior work~\cite{marcus_role_2021,ryan_now_2019}, we visually inspect our data in \autoref{fig:pta} and conclude that the parallel trends assumption is satisfied in all but two cases, daily post rate and daily removal rate for the gold outcome, which we exclude from the DiD analysis.  
This ensures the internal validity of the DiD models for the other outcomes and any observed effects~\cite{did_columbia_2016}.

We use our daily aggregates in the baseline and observation windows for each user in the following regression:
$$ y_{a, d} ~ \sim a_t*\delta(d \geq 0)  + d + s_a$$

where $y_{a, d}$ is the value of outcome $y$ for author $a$ at $d$ days after treatment/placebo, $a_t$ indicates whether author $a$ was treated, $d$ represents the number of days since treatment/placebo, $\delta(d \geq 0)$ indicates whether the aggregated data occurred after treatment, and $s_a$ is the stratum identifier of the author (see \autoref{matching}). We conduct one regression for each treatment and each outcome.

\subsubsection{Findings}

The results of our DiD analysis are reported in \autoref{tab:DiD}.
The increase column is calculated from the coefficient on the $a_t*\delta(d \geq 0)$ term which 
\edit{is the interaction between time and treatment group. This}
indicates the effect of treatment on treated users in the observation window \edit{compared to if they had not received the treatment, i.e., the counterfactual}.
To reduce the likelihood of spurious significant relationships, we perform a Bonferroni correction by multiplying the $p$-values of each coefficient by the product of the number of regression tests (\editprime{8}) and the number of independent variables per regression (4).
To capture the effect size, we also report the Cohen's $d$ measured between the difference in each dependent variable before and after treatment for each treated user compared to the same difference before and after placebo for each control user. 

We find that both forms of treatment lead to large positive increases in treated users' score on future posts compared to if they had not received the treatment. More specifically, receiving gold and being highly upvoted correlate with a user receiving 19.28\% and 57.11\% higher score than they would have, respectively. 
Both treatments have much more modest, but still significant, effects on the average number of gold received on future posts. The gold treatment encourages 1.63\% more gold in the future while the score treatment encourages an increase of less than 1\% more gold. 

\begin{table}[t]
\sffamily
    \centering
        \caption{Difference-in-differences regression analysis results. We report the percent increase observed in the dependent variable for each treated user compared to if they had never received the treatment. Regressions with Bonferroni-corrected $p$-values are reported with asterisks (* $p<0.05$, ** $p<0.01$, *** $p<0.001$).
        Cohen's $d$ reports the effect size of treatment on the dependent variables. We see significant increases in score and gold as a result of either treatment, and both \editprime{increases in text difference and} daily post rate and a decrease in daily removal rate from the score treatment.}
   \begin{tabular}{llr@{}lr}
\toprule
 \textbf{Treatment} & \textbf{DV}  & \textbf{Increase} & & \textbf{Cohen's $d$}   \\
\midrule
\multirow[t]{4}{*}{Gold} & Avg. Score & 19.28\% & *** & 0.16 \\
 & Avg. \# Gold & 1.63\% & *** & 0.27 \\
 & \editprime{Avg. Text Difference} & \editprime{-0.23\%} &  & \editprime{-0.05} \\
\\
\multirow[t]{4}{*}{Score} & Avg. Score & 57.11\% & *** & 0.14 \\
 & Avg. \# Gold & 0.2\% & *** & 0.04 \\
 
 & \editprime{Avg. Text Difference} & \editprime{0.34\%} & \editprime{***} & \editprime{0.01} \\
 & Daily Post Rate & 2.24\% & *** & 0.04 \\
 & Daily Removal Rate & -2.58\% & *** & -0.02 \\
\bottomrule
\end{tabular}
\Description{Percent increase in each dependent variable after treatment, split by the form of treatment. The final column reports the effect size measured by Cohen's d.}
    \label{tab:DiD} 
\end{table}

\editprime{Interestingly, users treated with score post content 0.34\% more distinct from their treated post than they would have if they had not received the treatment.  They also post 2.24\% more per day than they would have and  experience a negative effect on removal rate. This demonstrates that being highly upvoted might encourage users to post more frequent positive contributions while learning high-level community norms and thus violating them less often.}

Prior work on the platform Slashdot suggests that less popular posts may be more likely to be removed by volunteer moderators in a distributed moderation system~\cite{lampe_crowdsourcing_2014}. Therefore the decrease in removal rate after treatment may be attributable to the concurrent increase in score and not an indication of norm acquisition. However, other research on Reddit failed to confirm this trend of less harsh moderation decisions pertaining to higher-scored posts~\cite{vargo_deciding_2024}. 

Overall, receiving positive feedback of either kind encourages users to contribute posts that are better-received by their communities, and in some cases, also promotes user engagement and norm acquisition. 
Given the increase in future positive feedback and the decrease in norm-violations, we say that being highly upvoted encourages users to post higher-quality content according to both the community and the centralized moderation team.

\subsection{RQ1b: How long do the effects of positive feedback on user behavior last?}

\begin{figure*}
    \centering
    \includegraphics[width=.85\linewidth]{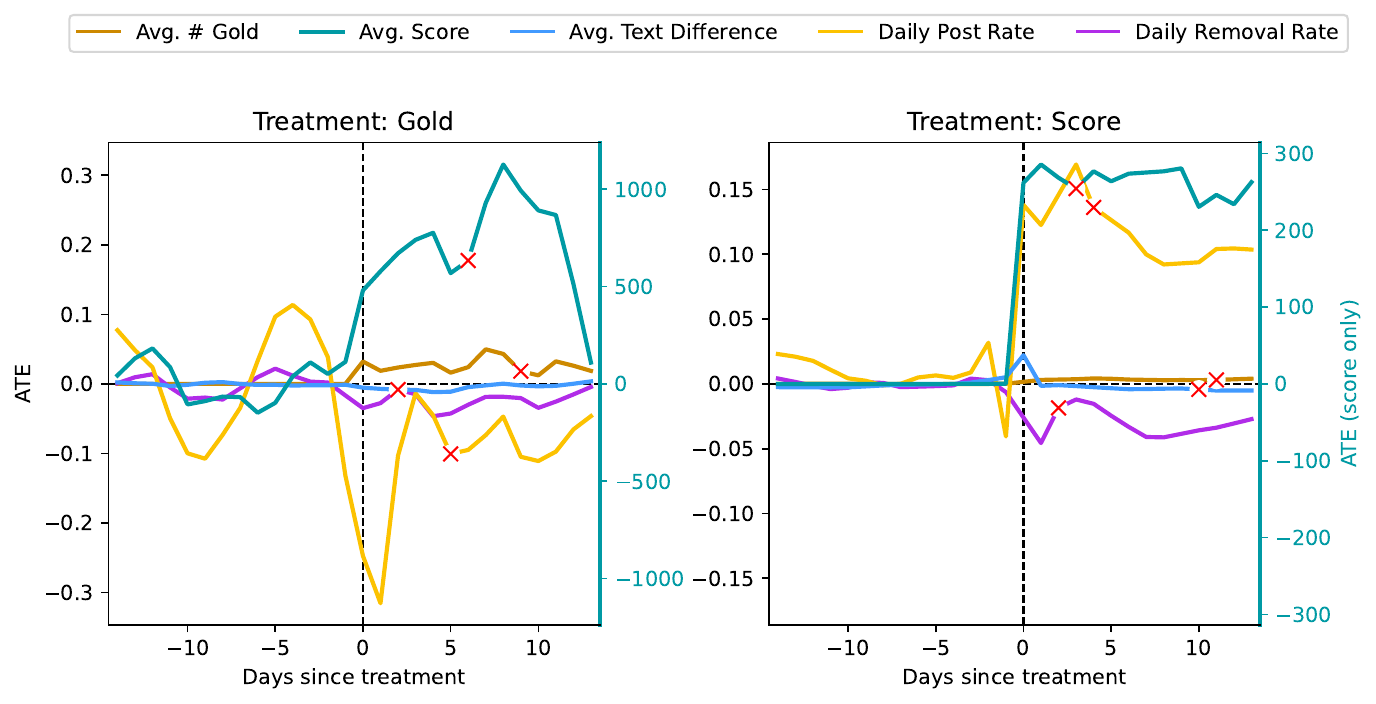}
    \caption{Average treatment effects (ATE) of each treatment on the \editprime{five} outcomes. The rightmost $y$-axes correspond to the ATE of the score outcome which has a much larger range. The ``x" markers indicate the days at which each outcome measure reaches saturation (i.e., decreases or stays the same for two consecutive days). The figure demonstrates the lasting effect of the score treatment on each outcome
    compared to the gold treatment, where the ATEs mostly return to zero within two weeks.}
    \Description{A plot with two subplots. Both plots show the average treatment effects for each outcome from the start of the baseline period to the end of the observation window. The leftmost plot shows these values for the gold treatment and the rightmost plot visualizes the daily ATE values for the score treatment. The 'x' icons on the leftmost plot show that the ATE for daily removal rate reaches saturation first, followed by daily post rate, average score, and finally average number of gold. Text difference does not rebound for the gold treatment. In the rightmost plot, removal rate reaches saturation first, then average score, daily post rate, text difference, and average number of gold.}
    \label{fig:ATE}
\end{figure*}

To determine how long the effect of each treatment lasts (RQ1b), we compute the Average Treatment Effect (ATE) for each of our \editprime{five} outcomes on each day in our study period. ATE is computed as the difference between the treated users' change in outcome and the control users' change in outcome  before and after treatment. We visualize the ATE values in \autoref{fig:ATE} averaged across each day in the baseline and observation windows. For each outcome, we also identify the day at which it reaches \textit{saturation}, meaning the ATE for the outcome stops increasing for two consecutive days after treatment. 
Formally, our saturation point is defined as the day $d$ such that $ATE(d) \leq ATE(d-1) \leq ATE(d-2)$. For the removal rate outcome, we reverse the definition of saturation point since we are interested in seeing decreases in norm violations. In other words, the saturation point is the first day $d$ for which $ATE(d) \geq ATE(d-1) \geq ATE(d-2)$. These points are denoted by red ``x" symbols in \autoref{fig:ATE}.

We observe that for the gold treatment, the ATE values for all \editprime{five} outcomes return to nearly zero by the end of the observation window, indicating minimal differences between treated and control users. For the users treated with score, however, the ATE of their score and post rate in the observation window remain well above zero at the end of the two-week observation window, even though both measures reach their saturation point by day 5. This indicates that the effect of being highly-upvoted on future upvotes and engagement may last longer than the effect of being gilded. The elevated scores in the observation window may also serve as additional treatments, creating a snowball effect.

\subsubsection{Rebound}

We extend this analysis by exploring how often the outcome variables return roughly to their baseline values. 
Each user's \textit{baseline region} for an outcome measure $\mathcal{M}$ is defined as the range $[0.9*\mathcal{M}_{\text{baseline}}, 1.1*\mathcal{M}_{\text{baseline}}]$. 
For each treatment-outcome pair, we find all users who return to their baseline region for two consecutive posts (i.e., rebound). 
\autoref{tab:rebound} summarizes the number of users who rebound for each outcome. The percentages in \autoref{tab:rebound} only consider users who post twice in their observation windows to satisfy the definition of rebound.

We notice that most treated users do not rebound on the score outcome within \obssize days, reflecting the trend shown in \autoref{fig:ATE} that score is elevated throughout the entire observation window. Daily post rate follows this same trend regardless of the treatment.
However, for the average number of gold, nearly all users in both treatments rebound to their baseline values within two weeks. This is understandable given the rarity of gilding in the dataset. 

Finally, we observe that the percentage of users who rebound to baseline \editprime{text difference} and removal rates after either treatment is much more evenly split. This suggests that the treatments \editprime{have a lasting impact on the content of some users' future posts and} teach some users about the norms more effectively than others. Our subsequent RQ2 analysis explores this trend in more depth.

\begin{table*}[t]
    \centering
        \caption{Summary table describing the number of treated users that rebound. By definition, users must post at least twice in the 2-week post-treatment window in order to rebound, which is true for 2,599 (60.2\%) users treated with gold and 90,993 (35.27\%) users treated with score. The table shows that users almost always rebound on their average number of gold, rarely rebound on their average score and daily post rate, and sometimes rebound on their \editprime{text difference and} daily removal rate.}
\begin{tabular}{llrrcc}
\toprule
\textsf{\textbf{Treatment}}&  \textsf{\textbf{Outcome}}& \textsf{\textbf{\# Rebound}} & \textsf{\textbf{\# Never Rebound}} \\
\midrule
\multirow[t]{4}{*}{\textsf{Gold}} & \textsf{Avg. Score} & \textsf{74 (1.72\%)} & \textsf{2,525 (58.53\%)} \\
 & \textsf{Avg. \# Gold} & \textsf{2,547 (59.04\%)} & \textsf{52 (1.21\%)} \\
 & \textsf{\editprime{Avg. Text Difference}} & \textsf{\editprime{1,768 (40.98\%)}} & \textsf{\editprime{831 (19.26\%)}} \\
 & \textsf{Daily Post Rate} & \textsf{159 (3.69\%)} & \textsf{2,440 (56.56\%)} \\
 & \textsf{Daily Removal Rate} & \textsf{1,282 (29.72\%)} & \textsf{1,317 (30.53\%)} \\
\\
\multirow[t]{4}{*}{\textsf{Score}} & \textsf{Avg. Score} & \textsf{3,640 (1.41\%)} & \textsf{87,353 (33.87\%)} \\
 & \textsf{Avg. \# Gold} & \textsf{90,139 (34.95\%)} & \textsf{854 (0.33\%)} \\
 & \textsf{\editprime{Avg. Text Difference}} & \textsf{\editprime{44,509 (17.26\%)}} & \textsf{\editprime{46,484 (18.02\%)}} \\
 & \textsf{Daily Post Rate} & \textsf{468 (0.18\%)} & \textsf{90,525 (35.10\%)} \\
 & \textsf{Daily Removal Rate} & \textsf{52,314 (20.28\%)} & \textsf{38,679 (15.00\%)} \\
\bottomrule
\end{tabular}
    \label{tab:rebound}
    \Description{Number of users who rebound and never rebound for each treatment-outcome pair. Percentages are reported in parentheses based on the number of each kind of treated user that posted twice in their observation window.}
\end{table*}

\section{RQ2: How do the effects of positive feedback vary across different types of recipients?}

Our RQ1 analysis raised questions about differential effects of treatment: do some users or communities benefit from treatment more than others?
We conduct two subsequent regression analyses to explore this question.

\subsection{RQ2a: What types of users experience stronger effects of receiving positive feedback?}

We compute Individual Treatment Effects (ITE) of each of our treatments on all \editprime{five} outcomes to evaluate user-level differential effects. ITE is used in prior work to compute how much a single treated user is affected by the treatment~\cite{saha_social_2019}. In our work specifically, ITE is computed as the difference between a treated user's change in behavior after treatment and the average change in all control users' behavior from the same stratum after placebo.
More precisely, for an individual treated user $u_{t,s}$ from stratum $s$, we compute the ITE for each outcome as follows: 
\begin{align*}
    \textit{ITE}_{\text{outcome}}(u_{t,s})=\  & \Delta u_{t,s}(\text{outcome}) \\
    &-  \text{mean}(\Delta u_{c,s}(\text{outcome}): u_{c,s} \in \mathcal{D}_{\textit{control,} s})
\end{align*}

where $\mathcal{D}_{\textit{control,} s}$ is the set of all control users in stratum $s$. Users who do not post in the observation window are excluded from this analysis since they do not have post-treatment data.

We then perform one linear regression analysis for each treatment-outcome pair using the outcome's ITE as the dependent variable. Baseline values for our \editprime{five} outcomes are independent variables along with the author's age, karma (i.e., the sum of scores on all posts) in the month prior to the baseline window, stratum, and the strength of treatment (i.e., number of gold or score on treated post). 
We take the log of each numerical independent variable.

\subsubsection{Findings}
\label{sec:ITE}
\autoref{tab:user_ITE} reports the coefficients from this regression. The bold coefficients are those we discuss in this section. All significance tests are Bonferroni-corrected based on the number of independent variables and regressions run. For the gold treatment, each $p$-value is multiplied by \editprime{40} (\editprime{5} regressions, \editprime{8} IVs) and for score, each $p$-value is multiplied by \editprime{45} (\editprime{5} regressions, \editprime{9} IVs).
This regression analysis demonstrates that the strength of either treatment may lead to stronger effects. For example, posts treated with more gold are positively correlated with the treated user experiencing stronger effects on their future number of gold. 
Similarly, users who are treated with higher score have more extreme effects on their future positive feedback than users with more modest treatments. With this score treatment, however, we see that being treated with lower score also correlates with a larger effect on removal rate. 

We also see trends aligned with the ideas of satiation and deprivation introduced by \citet{miltenberger_behavior_2016}. Specifically, our regression shows that users who have higher baseline average score experienced diminished effects of either treatment on their future score. 
Similarly, users with a higher average number of gold in their baseline window see less dramatic effects of the score treatment on their future number of gold. Both these trends reflect the idea that users are more strongly impacted by stimuli (e.g., gold or upvotes) if they have been deprived of that stimulus.

Additionally, the age of a user's account seems to affect the intensity of the impact of the score treatment on some of the outcomes. Specifically, 
newer accounts/authors experience the largest effects of being highly upvoted on their daily posting frequency and daily removal rates. Based on our results in \autoref{tab:DiD}, this means larger increases in post rate and larger decreases in norm violation rates for newer authors.
Thus, Reddit newcomers may be especially motivated by score to both continue contributing to a community and also learn more about the community's norms.

\editprime{Younger recipients of the score treatment also experience smaller increases in text difference, meaning newcomers post content in the future more similar to the post that received the reward compared to more established Reddit users. }

\begin{table*}
\footnotesize
\caption{Individual Treatment Effects (ITE) regression results reported with Bonferroni-corrected $p$-values (* $p<0.05$, ** $p<0.01$, *** $p<0.001$). 
Coefficients with $p>0.05$ are excluded. 
Cells that are grayed out indicate independent variables not present in the regression. \textbf{Bold} coefficients are discussed in \autoref{sec:ITE}.}
\resizebox{\linewidth}{!}{
    \begin{tabular}{l|r@{}lr@{}lr@{}lr@{}lr@{}l|r@{}lr@{}lr@{}lr@{}lr@{}l}
    \textsf{\textbf{Treatment}}	$\rightarrow$ & \multicolumn{10}{c}{\textsf{\textbf{Gold}}} & \multicolumn{10}{c}{\textsf{\textbf{Score}}} \\
    \cmidrule(lr){1-1}\cmidrule(lr){2-11}\cmidrule(lr){12-21}
\textsf{\textbf{IVs}} $\downarrow$ \textsf{\textbf{ITE}} $\rightarrow$ & \textsf{\textbf{Score}} & & \textsf{\textbf{\# Gold}} & & \textsf{\editprime{\textbf{Text Diff.}}} & & \textsf{\textbf{Post Rate}} & & \textsf{\textbf{Rem. Rate}} &  & \textsf{\textbf{Score}} && \textsf{\textbf{\# Gold}} && \textsf{\editprime{\textbf{Text Diff.}}} & & \textsf{\textbf{Post Rate}} & &  \textsf{\textbf{Rem. Rate}} &  \\
\midrule
\textsf{Author Age} &  &  &  &  &  &  &  &  &  &  &  &  &  &  & \textsf{\editprime{\textbf{4.6E-04}}} & \textsf{\editprime{\textbf{***}}} & \textsf{\textbf{-9.7E-03}} & \textsf{\textbf{***}} & \textsf{\textbf{-2.6E-03}} & \textsf{\textbf{***}} \\
\textsf{Baseline Avg. Score} & \textsf{\textbf{-5.5E-05}} & \textsf{\textbf{***}} & \textsf{1.8E-06} & \textsf{***} &  &  &  &  &  &  & \textsf{\textbf{-1.4E-02}} & \textsf{\textbf{***}} &  &  &  &  & \textsf{-3.1E-04} & \textsf{***} & \textsf{1.4E-03} & \textsf{***} \\
\textsf{\editprime{Baseline Text Diff.}} &  &  &  &  & \textsf{\editprime{-5.5E-01}} & \textsf{\editprime{***}} &  &  & \textsf{\editprime{2.0E-01}} & \textsf{\editprime{**}} & \textsf{\editprime{-1.5E-01}} & \textsf{\editprime{***}} & \textsf{\editprime{-5.9E-03}} & \textsf{\editprime{***}} & \textsf{\editprime{-8.0E-01}} & \textsf{\editprime{***}} & \textsf{\editprime{-3.2E-02}} & \textsf{\editprime{***}} & \textsf{\editprime{2.5E-01}} & \textsf{\editprime{***}} \\
\textsf{Baseline Post Rate} &  &  &  &  &  &  & \textsf{-2.3E-02} & \textsf{***} &  &  & \textsf{-8.8E-01} & \textsf{***} & \textsf{-4.2E-03} & \textsf{***} & \textsf{\editprime{-7.2E-03}} & \textsf{\editprime{***}} & \textsf{-2.0E-01} & \textsf{***} & \textsf{3.2E-01} & \textsf{***} \\
\textsf{Baseline Rem. Rate} &  &  &  &  &  &  & \textsf{-7.1E-02} & \textsf{*} & \textsf{-2.6E-01} & \textsf{***} & \textsf{2.2E+00} & \textsf{***} &  &  & \textsf{\editprime{4.4E-02}} & \textsf{\editprime{***}} & \textsf{-3.1E-01} & \textsf{***} & \textsf{-2.4E+00} & \textsf{***} \\
\textsf{Karma Last Month} & \textsf{1.1E-01} & \textsf{***} &  &  &  &  &  &  &  &  & \textsf{1.8E-02} & \textsf{***} &  &  & \textsf{\editprime{5.4E-04}} & \textsf{\editprime{***}} & \textsf{1.0E-03} & \textsf{*} &  &  \\
\textsf{Gold Received} &  &  & \textsf{\textbf{3.0E-02}} & \textsf{\textbf{**}} &  &  &  &  &  &  &  \cellcolor{gray} & \cellcolor{gray} &  \cellcolor{gray} & \cellcolor{gray} & \cellcolor{gray} & \cellcolor{gray} & \cellcolor{gray} & \cellcolor{gray} & \cellcolor{gray} & \cellcolor{gray}  \\
\textsf{Score Received} & \cellcolor{gray} & \cellcolor{gray} &  \cellcolor{gray} & \cellcolor{gray} & \cellcolor{gray} & \cellcolor{gray} & \cellcolor{gray} & \cellcolor{gray} & \cellcolor{gray} & \cellcolor{gray} &\textsf{\textbf{ 9.7E-02}} & \textsf{\textbf{***}} & \textsf{\textbf{5.7E-04}} & \textsf{\textbf{***}} &  &  &  &  & \textsf{\textbf{-3.0E-03}} & \textsf{\textbf{***}} \\
\textsf{Baseline Avg. \# Gold} & \cellcolor{gray} & \cellcolor{gray} &  \cellcolor{gray} & \cellcolor{gray} & \cellcolor{gray} & \cellcolor{gray} & \cellcolor{gray} & \cellcolor{gray} & \cellcolor{gray} & \cellcolor{gray} &  &  & \textsf{\textbf{-5.0E-01}} & \textsf{\textbf{***}} &  &  &  &  &  &  \\
\\
\textsf{R-squared} & \textsf{0.111} &  & \textsf{0.052} &  & \textsf{\editprime{0.395}} &  & \textsf{0.109} &  & \textsf{0.080} &  & \textsf{0.108} &  & \textsf{0.183} &  & \textsf{\editprime{0.579}} &  & \textsf{0.058} &  & \textsf{0.250} &  \\
\bottomrule
\end{tabular}}
    \label{tab:user_ITE}
    \Description{Regression results for the ten regressions measuring the effect of the independent variables on the ITE values for each treatment-outcome pair.}
\end{table*}

\subsection{RQ2b: What types of communities experience stronger average effects of receiving positive feedback?}

We follow up this user-level analysis with a community-level analysis involving Community Treatment Effects (CTE) for each treatment and outcome pairing. 
We define the CTE of an outcome for a Reddit community $r$ as follows: 
\begin{align*}
    \textit{CTE}_{\text{outcome}}(r) =\  &\text{mean}(\Delta u_{t,r}(\text{outcome}): u_{t,r} \in \mathcal{D}_{\textit{treated,} r})\\ 
    &-  \text{mean}(\Delta u_{c,r}(\text{outcome}): u_{c,r} \in \mathcal{D}_{\textit{control,} r})
\end{align*}

where $\mathcal{D}_{\textit{treated,} r}$ and $\mathcal{D}_{\textit{control,} r}$ represent the sets of all treated and control users in subreddit $r$ respectively.

Similar to the previous ITE analysis, we use our CTE values as dependent variables for ten linear regression analyses with subreddit-level monthly activity statistics from our entire study period as independent variables. This includes post rate, average number of gold, average score, the number of unique authors, and the number of subscribers.

\subsubsection{Findings} After Bonferroni correction, these regressions yielded only two significant coefficients revealing that subreddits that gild more often see larger effects of the score treatment on their future number of gold and text difference. The lack of significant results indicates that the treatments have similar effects regardless of the community in which they occur.
\section{Discussion}

Our work contributes a computational approach and a causal inference framework through which researchers can examine the differential effects of treatment outcomes. We outline a method of identifying treatment and control posts, performing a difference-in-differences analysis, and determining user- and community-level factors that influence the strength of the treatment effect.
This methodology can be adapted with different treatments, outcomes, and platforms to fit researchers' needs. Additionally, while we focused on two types of positive feedback available on Reddit, upvotes in particular are almost universally applicable across social media platforms and have analogous features on Facebook, Instagram, X, Stack Overflow, and YouTube, among others. This ubiquity of upvotes enhances the generalizability of our work. Researchers interested in other platforms can use these findings to inform research into analogous features\edit{, and other forms of formal feedback,} and extend this work with analyses of positive feedback signals not available on Reddit.

Our findings demonstrate the power Reddit affords community members through signals of positive feedback on the interface. 
We also revealed the differential effects of positive feedback, specifically that different forms of feedback have different effects on different types of recipients. 
These results have important implications for content moderation and motivate the need to emphasize positive feedback in HCI research and in platform design.

\subsection{Prioritize highlighting contributions from users recently deprived of positive feedback.}

A key differential effect we observed relates to the principle of deprivation in positive reinforcement theory. The principle states that a stimulus has larger impacts on subjects deprived of it~\cite{miltenberger_behavior_2016}. 
Our findings confirm this theory in the context of positive feedback on Reddit, highlighting that users who had not recently received gold experienced stronger effects of being gilded on their future gold received. Similarly, users who had not recently been highly-upvoted experienced larger increases in their future score when experiencing the score treatment. 

\subsubsection*{Design recommendations.}
\textit{We recommend that designers and HCI researchers consider the principles of satiation and deprivation to best utilize valuable community resources like time and effort.}
Specifically, we suggest that they \edit{enable moderators and community members to maximize the potential effects of their positive feedback by building on work highlighting users' history of toxicity~\cite{im_synthesized_2020} and instead providing signals of how deprived or satiated an author is, or by prioritizing content in social media feeds contributed by authors deprived of positive feedback. } 

\subsection{Provide mechanisms to bring attention to high-quality newcomer contributions.}

As moderators are largely responsible for sifting through norm-violating content, they stand to benefit the most from users learning to contribute norm-abiding content. However, the idea of norm acquisition can be challenging, especially for newcomers to a community~\cite{kraut_building_2011}, and is often a barrier to entry when users want to contribute to a community in good faith~\cite{schaumberg_shame_2022,kashima_acquisition_2013,rajadesingan_quick_2020}. 
Some platforms (e.g., Wikipedia) have developed systems to assist newcomers in learning to make quality contributions, but research has shown that some of these methods may be hurting newcomers more than they are helping~\cite{schneider_accept_2014}.
Additionally, while some norms are consistent platform-wide, there are other norms that may only apply to a handful of subreddits~\cite{chandrasekharan_internets_2018}. Thus, even established Reddit users who participate in many communities can face difficulties trying to comply with the rules of other communities~\cite{jhaver_did_2019}. 

We showed that positive feedback in the form of upvotes results in the recipient posting fewer norm-violating posts than they would have without the treatment, implying that recipients are learning community norms. 
Additionally, newer accounts see stronger effects on their norm-adherence after treatment compared to more established Reddit users indicating that platform-level newcomers benefit most from positive feedback in the context of norm acquisition. \editprime{Newcomers also more closely match the text of their future posts to the one that was rewarded, suggesting the power of positive feedback to shape the actual content of newcomer contributions.}

\subsubsection*{Design recommendations.}
These findings indicate that \textit{positive feedback as a means to support norm acquisition is imperative for the success of moderation teams}. 
Specifically, we know that exposure to harmful content is concerning to volunteer moderators~\cite{steiger_psychological_2021}. Our work shows that users who receive positive feedback are less likely to be removed and thus not exposing moderators to potentially harmful, norm-violating content as frequently as they would have. Therefore, the use of positive feedback may have positive effects on moderator well-being.

To enable this norm acquisition, we call on designers and HCI researchers to explore explicit ways for moderation teams to encourage established community-members to review newcomer contributions and provide positive feedback when applicable. 
\edit{For example, this may be a highlighting mechanism similar to YouTube creator hearts which have been shown to increase positive engagement for hearted comments~\cite{choi_creator_2024}. 
Designers may also consider building off a prior intervention that nudges users to preemptively reflect on their participation~\cite{chang_thread_2022} and build a system to nudge users to specifically engage with newcomer content. 
Prior work on visualizing conversational metrics~\cite{choi_convex_2023} can be extended to develop tools that guide moderator attention towards desirable behavior.
Finally, platforms can incorporate} a dedicated feed of newcomer contributions within communities \edit{to enable easy discovery of newcomer content.}

\subsection{Platforms need positive feedback to sustain user motivation to contribute.}
\citet{kraut_building_2011} state that ``to be successful, online communities need the people who participate in them to contribute the resources on which the group’s existence is built.'' In the context of Reddit, those resources include memes, links to news articles, or other subreddit-specific contributions. Without users making such contributions, platforms would decline in popularity and lose out on profit. 
Thus, platforms must maintain user motivation.

User motivation can be intrinsic or extrinsic~\cite{kraut_building_2011}. In the context of our research, intrinsically-motivated users contribute high-quality content because they enjoy doing so. \citet{kraut_building_2011} highlight performance-based positive feedback as an especially effective method of enhancing users' intrinsic motivation. Therefore, users who receive community-wide approval through positive feedback may experience increased intrinsic motivation to participate. 

Extrinsically-motivated users may be more interested in making high-quality posts as a means to increase their reputation (i.e., karma) in the community. Rewards such as gold and upvotes are extrinsic motivators in themselves, which can help these types of users feel more motivated to participate~\cite{kraut_building_2011}.

\subsubsection*{Implications for platforms and moderators.}
Regardless of whether our treated users' motivation was intrinsic or extrinsic, it was enhanced by positive feedback~\cite{kraut_building_2011}. 
Our findings confirm that users who contribute posts that get highly-upvoted increase their posting rate, indicating a potential increase in motivation to participate. Consequently, we assert that \textit{positive feedback is crucial to platform success given its significant role in supporting user motivation}.

\subsection{Centralized and distributed moderation systems can support communities through positive feedback.}

Reddit moderation consists of a centralized moderation team and a community of users implicitly engaging in distributed moderation practices. Based on our findings, both facets of Reddit's moderation system can help improve the quality of content and norm-adherence in communities.

\subsubsection*{Users already have the power to shape their communities through distributed moderation.}

The sorting algorithms of feeds on platforms like Reddit are often opaque to users and referred to as ``black boxes'' since platforms do not disclose the metrics they use to generate these feeds~\cite{chan_understanding_2024}. However, users may have hypotheses about how the sorting works~\cite{eslami_i_2015}. For example, surveyed moderators reported providing various forms of positive feedback to boost the visibility of content, implying that signals like score and gold affect a post's position in an algorithmically-curated feed~\cite{lambert_positive_2024}.
However, our research has shown that positive feedback mechanisms can go beyond influencing the sorting algorithm and actually impact the quality of content generated by users in the future. 

Because gold and upvotes both encourage users to contribute posts that receive higher score and more gold, we conclude that both forms of positive feedback encourage
users to contribute posts that will be better received by the community. 
This indicates that
the community overall is more satisfied with contributions made by recipients of positive feedback. 
This reveals  the power communities already have to proactively encourage a better pool of content through positive reinforcement.
Thus, though moderators have additional weight and power to make decisions and enforce norms in a community, they are not the only ones capable of effecting positive change. 

Though this work focuses on the capacity of community-members to facilitate distributed moderation through existing interface signals, centralized moderation teams on Reddit can learn from our analysis to inform how they moderate. 
Prior work found that YouTube comments receiving a ``heart'' from the creator of a video resulted in increased positive feedback from the community as a whole~\cite{choi_creator_2024}. 
Thus, when moderators demonstrate their approval of a contribution through highlighting, the community at large may reward it with positive feedback mechanisms that we found to positively impact user-level outcomes.

\subsubsection*{Implications for online moderation.}
We found that users already have the power to shape their communities through distributed moderation. When members of a community collectively reward a contribution, we see a significant positive impact of the distributed effort. Thus, \textit{platforms and researchers should support community-driven moderation through positive feedback to give communities the agency to proactively curate their content}.

Moreover, members of centralized moderation teams can curate posts to be rewarded through distributed moderation.
Since Reddit does not explicitly provide moderator-specific feedback mechanisms, \textit{we suggest moderation teams focus on influencing the visibility of posts which can impact the amount of positive feedback they receive.} 
Moderators can pin high-quality posts or periodically consolidate posts in a moderator-curated list in the subreddit sidebar, both of which moderators have reported doing~\cite{lambert_positive_2024}. 

\subsection{Limitations and Future Work}

Our research opens many avenues for the exploration of positive reinforcement and quantifying desirable content in online spaces. We summarize the limitations of our work and propose extensions to address them.

\subsubsection{Build models of desirability.} 
Our analysis does not consider
\edit{whether the posts that received the positive feedback are deserving of such approval. Definitions of desirability can vary greatly~\cite{lambert_investigating_2024, goyal_uncovering_2024}, thus future work should explore what is considered desirable within and across online communities, and develop context-sensitive approaches to model and identify desirable behavior online}.

\subsubsection{\editprime{Extend understanding of content-specific reinforcement.}} Our analysis showed that recipients of positive feedback \editprime{receive even more positive feedback on subsequent posts, but that the content of those posts is more distinct from the rewarded post than it would have been without the treatment. This suggests that receiving positive feedback helps users apply community norms beyond making redundant contributions. However, this effect was small and it is possible that working with posts, which do not always have much text, does not allow us to capture content-level reinforcement. We suggest researchers extend this work by exploring the effects of positive feedback on reinforcing specific types of content within both posts and comments.}

\subsubsection{Explore the effects of positive feedback on bystanders.}
While the effects studied in this research were all specific to the recipients of positive feedback, there may also be effects on users who interact with rewarded content through vicarious reinforcement. \citet{jhaver_bystanders_2024} show that bystanders witnessing post removal explanations experience effects on their future behavior. Future work is needed to understand whether bystanders of positive feedback are similarly affected and whether users can learn community norms by witnessing such positive interventions.

\subsubsection{Compare community feedback to moderator-specific forms of feedback}

Our choice of treatments is limited in that some moderators do not consider upvotes a signal of quality~\cite{jiang_trade-off-centered_2023}. This can be problematic in communities that require expertise for participation, since users giving out upvotes may not be equipped to evaluate how much a post deserves positive feedback. 
In such communities, moderators may be better judges of quality.
Prior work shows that moderators are utilizing moderator-specific signals as a means for positive reinforcement~\cite{lambert_positive_2024}, thus future work should study how these forms of feedback affect the recipients and whether positive feedback from a position of authority has more or less capacity to reinforce user-level outcomes than anonymous forms of feedback. 
\section{Conclusion}

In this work, we examined the effects of two forms of positive feedback on Reddit users and study whether positive reinforcement can be employed through distributed moderation practices. We found that both receiving gold on a post and being highly-upvoted by the community encourage users to post content in the future that receives even more positive feedback than they would have otherwise. Through a user-specific analysis, we also find that receiving larger amounts of positive feedback is associated with larger effects on future positive feedback though users already satiated by recent positive feedback will not see as strong effects. Importantly, we also found that newcomers experience stronger effects of being highly-upvoted with respect to their future rate of contribution and norm acquisition. Based on our findings, we contribute design recommendations and discuss the implications of our work with respect to moderators, communities, and platforms beyond Reddit.

%
\begin{acks}
We thank the members of the Social Computing Lab (SCUBA) at the University of Illinois Urbana-Champaign for their feedback and input on this work. We also thank the anonymous reviewers
for their thoughtful and valuable reviews.

\end{acks}

\bibliographystyle{ACM-Reference-Format}
\bibliography{bib}


\end{document}